\documentclass{emulateapj}
\usepackage{apjfonts}
\usepackage{rotating}

\lefthead{YEE ET AL.} 
\righthead{MASSIVE GALACTIC DISK PLANET}

\newcommand{\bdv}[1]{\mbox{\boldmath$#1$}}

\def\au{{\rm AU}} 
 
\def\kms{{\rm km}\,{\rm s}^{-1}}
\def\masyr{{\rm mas}\,{\rm yr}^{-1}}
\def\kpc{{\rm kpc}}

\def\max{{\rm max}}
\def\rel{{\rm rel}}

\def\eff{{\rm eff}}
\def\hel{{\rm hel}}
\def\geo{{\rm geo}}
\def\e{{\rm E}}
\def\bpi{{\bdv\pi}}
\def\bmu{{\bdv\mu}}

\begin{document}
\title{MOA-2013-BLG-220Lb: Massive Planetary Companion to Galactic-Disk Host}

\author{
J.~C.~Yee$^{U1,U2,U14\dag}$,
C.~Han$^{U3,\dag,\spadesuit}$, 
A.~Gould$^{U1,\dag}$, 
J.~Skowron$^{O1,\diamondsuit}$,
I.~A.~Bond$^{M1,\ddag}$,
A.~Udalski$^{O1,\diamondsuit}$,
M. Hundertmark$^{R1,\clubsuit}$,
L.~A.~G.~Monard$^{U10,\dag}$,
I.~Porritt$^{U12,\dag}$,
P.~Nelson$^{U11,\dag}$,
V.~Bozza$^{N1,N2}$,\\
and\\
M.~D.~Albrow$^{U6}$,
J.-Y.~Choi$^{U3}$,
G.~W.~Christie$^{U4}$,
D.~L.~DePoy$^{U7}$,
B.~S.~Gaudi$^{U1}$,
K.-H.~Hwang$^{U3}$,
Y.~K.~Jung$^{U3}$,
C.-U.~Lee$^{U8}$,
J.~McCormick$^{U9}$,
T.~Natusch$^{U4,U13}$,
H.~Ngan$^{U4}$,
H.~Park$^{U3}$,
R.~W.~Pogge$^{U1}$,
I.-G.~Shin$^{U3}$,
T.-G.~Tan$^{U5}$\\
(The $\mu$FUN Collaboration),\\
F.~Abe$^{M2}$,
D.~P.~Bennett$^{M3}$,
C.~S.~Botzler$^{M5}$,
M.~Freeman$^{M5}$, 
A.~Fukui$^{M6}$, 
D.~Fukunaga$^{M2}$,
Y.~Itow$^{M2}$,
N.~Koshimoto$^{M7}$,
P.~Larsen$^{M8}$, 
C.~H.~Ling$^{M1}$, 
K.~Masuda$^{M2}$, 
Y.~Matsubara$^{M2}$,
Y.~Muraki$^{M2}$,
S.~Namba$^{M7}$, 
K.~Ohnishi$^{M9}$, 
L.~Philpott$^{M10}$,
N.~J.~Rattenbury$^{M5}$,
To.~Saito$^{M11}$,
D.~J.~Sullivan$^{M12}$,
T.~Sumi$^{M7}$,
W.~L.~Sweatman$^{M1}$, 
D.~Suzuki$^{M7}$,
P.~J.~Tristram$^{M13}$,
N.~Tsurumi$^{M2}$, 
K.~Wada$^{M7}$,
N.~Yamai$^{M14}$, 
P.~C.~M.~Yock$^{M5}$,
A.~Yonehara$^{M14}$ \\
(MOA Collaboration), \\
M.K.~Szyma{\'n}ski$^{O1}$,
K.~Ulaczyk$^{O1}$,
S. Koz{\l}owski$^{O1}$, 
R.~Poleski$^{U1,O1}$,
{\L}.~Wyrzykowski$^{O1,O2}$,
M.~Kubiak$^{O1}$,
P.~Pietrukowicz$^{O1}$,
G.~Pietrzy{\'n}ski$^{O1,O3}$,
I.~Soszy{\'n}ski$^{O1}$ \\
(OGLE Collaboration), \\
D. M. Bramich$^{R2}$, 
P. Browne$^{R1}$, 
R. Figuera Jaimes$^{R1,R4}$
K. Horne$^{R1}$,
S. Ipatov$^{R3}$, 
N. Kains$^{R4}$, 
C. Snodgrass$^{R5}$, 
I. A. Steele$^{R6}$,
R. Street$^{R7}$, \\
Y. Tsapras$^{R7,R8}$ \\
(The RoboNet Collaboration)\\
}

\affil{$^{U1}$Department of Astronomy, Ohio State University, 140 West 18th Avenue, Columbus, OH 43210, USA}
\affil{$^{U2}$Harvard-Smithsonian Center for Astrophysics, 60 Garden St., Cambridge, MA 02138, USA}
\affil{$^{U3}$Department of Physics, Chungbuk National University, Cheongju 361-763, Korea}
\affil{$^{U4}$Auckland Observatory, Auckland, New Zealand}
\affil{$^{U5}$Perth Exoplanet Survey Telescope, Perth, Australia}
\affil{$^{U6}$Department of Physics and Astronomy, University of Canterbury, Private Bag 4800, Christchurch 8020, New Zealand}
\affil{$^{U7}$Department of Physics and Astronomy, Texas A\&M University, College Station, Texas 77843-4242, USA}
\affil{$^{U8}$Korea Astronomy and Space Science Institute, 776 Daedukdae-ro, Yuseong-gu, Daejeon 305-348, Republic of Korea}
\affil{$^{U9}$Farm Cove Observatory, Centre for Backyard Astrophysics, Pakuranga, Auckland, New Zealand}
\affil{$^{U10}$Klein Karoo Observatory, Centre for Backyard Astrophysics, Calitzdorp, South Africa}
\affil{$^{U11}$Ellinbank Observatory, Ellinbank, Victoria, Australia}
\affil{$^{U12}$Turitea Observatory, Palmerston North, New Zealand}
\affil{$^{U13}$ Institute for Radio Astronomy and Space Research, AUT University, Auckland New Zealand}
\affil{$^{U14}$ Sagan Fellow}
\affil{$^{M1}$Institute of Information and Mathematical Sciences, Massey University, Private Bag 102-904, North Shore Mail Centre, Auckland, New Zealand}
\affil{$^{M2}$Solar-Terrestrial Environment Laboratory, Nagoya University, Nagoya, 464-8601, Japan} 
\affil{$^{M3}$University of Notre Dame, Department of Physics, 225 Nieuwland Science Hall, Notre Dame, IN 46556-5670, USA}
\affil{$^{M5}$Department of Physics, University of Auckland, Private Bag 92-019, Auckland 1001, New Zealand}
\affil{$^{M6}$Okayama Astrophysical Observatory, National Astronomical Observatory of Japan, Asakuchi, Okayama 719-0232, Japan}
\affil{$^{M7}$Department of Earth and Space Science, Osaka University, Osaka 560-0043, Japan}
\affil{$^{M8}$Institute of Astronomy, University of Cambridge, Madingley Road, Cambridge CB3 0HA, UK}
\affil{$^{M9}$Nagano National College of Technology, Nagano 381-8550, Japan}
\affil{$^{M10}$Department of Physics and Astronomy, The University of British Columbia, 6224 Agricultural Road Vancouver, BC V6T 1Z1, Canada}
\affil{$^{M11}$Tokyo Metropolitan College of Aeronautics, Tokyo 116-8523, Japan}
\affil{$^{M12}$School of Chemical and Physical Sciences, Victoria University, Wellington, New Zealand}
\affil{$^{M13}$Mt. John University Observatory, P.O. Box 56, Lake Tekapo 8770, New Zealand}
\affil{$^{M14}$Department of Physics, Faculty of Science, Kyoto Sangyo University, 603-8555, Kyoto, Japan}
\affil{$^{O1}$Warsaw University Observatory, Al. Ujazdowskie 4, 00-478 Warszawa, Poland}
\affil{$^{O2}$Institute of Astronomy, University of Cambridge, Madingley Road, Cambridge CB3 0HA, UK}
\affil{$^{O3}$Universidad de Concepci{\'o}n, Departamento de Astronomia, Casilla 160--C, Concepci{\'o}n, Chile}
\affil{$^{R1}$SUPA, School of Physics \& Astronomy, University of St Andrews, North Haugh, St Andrews KY16 9SS, UK}
\affil{$^{R2}$Qatar Environment and Energy Research Institute, Qatar Foundation, Tornado Tower, Floor 19, P.O. Box 5825, Doha, Qatar}
\affil{$^{R3}$Qatar Foundation, P.O. Box 5825, Doha, Qatar}
\affil{$^{R4}$European Southern Observatory, Karl-Schwarzschild-Str. 2, 85748 Garching bei M\"unchen, Germany}
\affil{$^{R5}$Max Planck Institute for Solar System Research, Max-Planck-Str. 2, 37191 Katlenburg-Lindau, Germany}
\affil{$^{R6}$Astrophysics Research Institute, Liverpool John Moores University, Liverpool CH41 1LD, UK}
\affil{$^{R7}$Las Cumbres Observatory Global Telescope Network, 6740 Cortona Drive, suite 102, Goleta, CA 93117, USA}
\affil{$^{R8}$School of Physics and Astronomy, Queen Mary University of London, Mile End Road, London E1 4NS, UK}
\affil{$^{N1}$Dipartimento di Fisica ``E. R. Caianiello,'' Universit\`a degli Studi di Salerno, Via S. Allende, I-84081 Baronissi (SA), Italy}
\affil{$^{N2}$Instituto Nazionale di Fisica Nucleare, Sezione di Napoli, Italy}
\affil{$^{\dag}$The $\mu$FUN Collaboration}
\affil{$^{\ddag}$The MOA Collaboration}
\affil{$^{\diamondsuit}$The OGLE Collaboration}
\affil{$^{\clubsuit}$The RoboNet Collaboration}
\affil{$^{\spadesuit}$Corresponding author}

\begin{abstract}
We report the discovery of MOA-2013-BLG-220Lb, which has a super-Jupiter
mass ratio $q=3.01\pm 0.02\times 10^{-3}$ relative to its host.  The
proper motion, $\mu=12.5\pm 1\,\masyr$, is one of the highest for
microlensing planets yet discovered, implying that it will be possible
to separately resolve the host within $\sim 7$ years.  Two separate lines
of evidence imply that the planet and host are in the Galactic disk.
The planet could have been detected and characterized purely with 
follow-up data, which has important implications for microlensing 
surveys, both current and into the LSST era.
\end{abstract}


\section{Introduction}

Because microlensing planet detections are based on observations
of a background source that is lensed by the planetary system,
rather than observations of the planetary system itself,
microlensing is unique in its ability to detect planets orbiting
extremely dim or dark hosts or even planets without hosts \citep{sumi11}.
For the same reason, however, microlensing planet hosts are often
difficult to characterize.  This can in principle be done by
simultaneously measuring two higher-order effects during the event,
yielding the Einstein radius $\theta_\e$ and the ``microlens parallax'' $\pi_\e$.
Then the lens mass $M$ and lens-source relative parallax $\pi_\rel$
are given by \citep{gould92}
\begin{equation}
M = {\theta_\e\over \kappa \pi_\e};
\qquad
\pi_\rel = \pi_\e\theta_\e;
\qquad
\kappa\equiv {4G\over c^2\au}\simeq 8.1\,{{\rm mas}\over M_\odot}.
\label{eqn:mpirel}
\end{equation}
This has been successfully carried out for a significant minority
of microlensing planets to date, and actually verified in one case
by direct imaging \citep{ob06109,ob06109b}.  However, while $\theta_\e$
has been measured in the great majority of published planetary events, 
$\pi_\e$ usually proves too difficult to measure.  In this
case, one only has the mass-distance constraint,
\begin{equation}
M \pi_\rel = {\theta_\e^2\over \kappa}.
\label{eqn:mpirel2}
\end{equation}

An alternate approach is to directly observe the lens in high-resolution
images, either under the ``glare'' of the source while they are still
superposed, provided that the lens is sufficiently bright \citep{mb11293B},
or by waiting for the lens and source to separate 
(Batista et al.\ 2014b, in preparation;  
Bennett et al.\ 2014, in preparation).  
Of course, direct imaging is ill-suited to detecting
dark hosts, but it can at least verify that they are dark, particularly
if the mass-distance constraint (Equation~(\ref{eqn:mpirel2})) is available.

Here we present MOA-2013-BLG-220Lb, with super-Jupiter
planet/host mass ratio $q=3.0\times 10^{-3}$.
Although the host (and so planet) mass is presently unknown, we show that
it is moving rapidly away from
the source ($\mu_\rel=12.5\pm 1\,\masyr$) and so can be imaged separately
within $\sim 7\,$yrs.

\section{Observations}

On 2013 Apr 1, UT 19:08, the Microlensing Observations in Astrophysics
(MOA) collaboration\footnote{https://it019909.massey.ac.nz/moa/} 
issued an alert that MOA-2013-BLG-220 was an
ongoing microlensing event at
(RA,Dec)$=(18:03:56.5, -29:32:41)$,  $(l,b)=(1.50,-3.76)$, based on
data taken in broad $RI$ band using their 1.8m telescope at Mt.\ John,
New Zealand.
$\mu$FUN issued its own alert 46 hours 
later saying that this was likely
to be a high-magnification event and thus very sensitive to planets
\citep{griest98}, which triggered observations by 
the Kleinkaroo Observatory.  
These data showed a sharp increase in brightness,
which was interpreted as evidence of a very high magnification
event and so triggered a further alert, but was actually due to the
onset of the anomaly (see Fig.~1).  This alert, issued as the event was
rising over Chile, noted that $\mu$FUN's own telescope at that location
was non-operational due to equipment problems, and made a particular
request that other telescopes at this longitude observe the event.

\begin{figure}
\epsscale{1.1}
\plotone{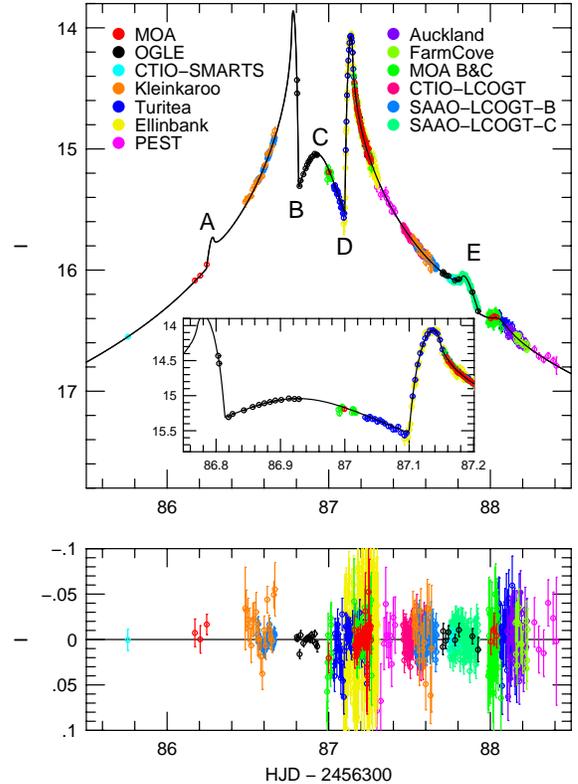}
\caption{\label{fig:one}
Best fit model and residuals for MOA-2013-BLG-220.  From the
lightcurve alone, it is clear that there are two caustic
crossings (ending at $\sim 86.8$ (B) and starting at $\sim 87.1$ (D)) 
and a cusp approach ($\sim 86.95$ (C)),
hence three cusps on one side of the primary lens, implying a planetary
mass ratio for the companion.  The fact that the time ($\sim 0.15\,$day) 
from the cusp approach (C) to second entrance (D)
is much shorter than
the time ($\sim 0.85\,$day) to the second exit (E) shows
that the caustic has a resonant topology, implying $|s-1|\ll 1$.
See Figure~\ref{fig:two}.
The near symmetry of the lightcurve about the cusp approach implies
that the source passed nearly perpendicular to planet-star axis.
All these predictions are confirmed by detailed modeling.  
See Table~\ref{tab:one}.
Kleinkaroo and PEST data are binned for plotting purposes but are
not binned in the light-curve fitting.}
\end{figure}

The Optical Gravitational Lens Experiment (OGLE) responded by
putting their 1.3m telescope at Las Campanas Chile, which is usually
dedicated to microlensing survey operations, into ``followup mode''.
Specifically, OGLE found that MOA-2013-BLG-220 lay in a gap between
OGLE mosaic-camera CCD chips in the template image of this field.  
Thus, although they did have occasional
observations of this target when the telescope pointing drifted slightly,
they were not monitoring these observations in real time and so did not
issue an alert.  However, in response to $\mu$FUN's alert, OGLE
both slightly changed the pointing of their telescope for this survey
field, and also dramatically increased the cadence.  It is not
uncommon for OGLE to increase cadence in response to interesting events
(e.g., \citealt{mb11293}), but altered pointing is much rarer.

The OGLE followup data showed a clear caustic exit, which was the
first unambiguous evidence for a planetary (or possibly binary) event.
This triggered a further alert.  In the end, followup data were taken
by seven $\mu$FUN telescopes, three RoboNet telescopes, one followup
telescope on the same site as the MOA survey telescope, and OGLE in 
follow-up mode.

The seven $\mu$FUN sites are
Kleinkaroo Observatory (30cm unfiltered, Calitzdorp, South Africa), 
Turitea Observatory  (36cm $R$-band, Palmerston, New Zealand),
Ellinbank Observatory (32cm $V$-band, Victoria, Australia),
PEST Observatory (30cm, unfiltered, Perth, Australia),
Auckland Observatory (40cm $R$-band, Auckland, New Zealand),
Farm Cove Observatory (36cm unfiltered, Pakuranga, New Zealand), and
CTIO SMARTS (1.3m $I$, $V$ bands La Serena, Chile), with all
but the last being amateur observatories.

All three RoboNet telescopes are 1m and robotic, with one 
at La Serena, Chile ($I$ band, CTIO-LCOGT) and two at Sutherland, South Africa
($I$ band, SAAO-LCOGT-B,C).  In particular, the data from Chile covered the
second caustic exit extremely well.

The MOA 61cm B\&C telescope ($I$ band, Mt.\ John, NZ) responded to the alert by 
obtaining intensive observations over the next two nights, in particular
tracing a subtle structure of the second caustic exit.

All data were reduced using difference image analysis 
\citep{alard98,bramich08}.

\begin{figure}[t]
\epsscale{1.1}
\plotone{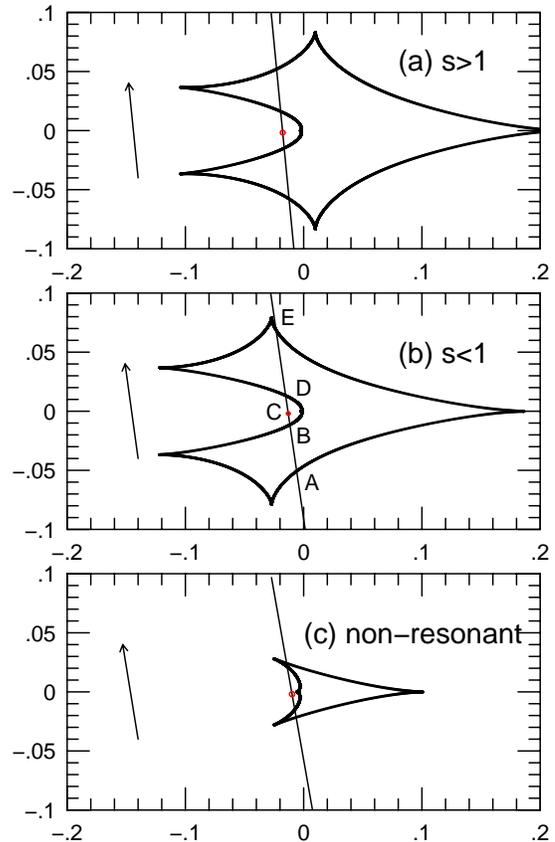}
\caption{\label{fig:two}
Three possible central caustic topologies that could be consistent
with lightcurve in Figure~\ref{fig:one}; a (b): resonant 6-sided caustic
with $s>1$ ($s<1$), and c: non-resonant 4-sided caustic (same for both
$s>1$  and $s<1$).  Trajectory must pass roughly perpendicular to
caustic axis, close to central cusp (as shown) in order to reproduce
the lightcurve's approximate symmetry.  The fact that the time
from the cusp approach (C) to the second entrance (D)
is much shorter than from cusp approach (C) to second exit (E)
rules out topology (c).
Topology (b) is preferred over (a) by $\Delta\chi^2\sim 6000$.
This preference is not easily traceable to gross features of the lightcurve
but see Section 3 for qualitative reasoning.
The caustic geometries in (a) and (b) correspond to the actual
best-fit solutions for the indicated topologies.}
\end{figure}

\section{Lightcurve Analysis}

\subsection{Topological Analysis}

Figure 1 shows the data together with the best-fit model.  
However,
the basic character of the event can be understood without reference
to any model.  The first feature is that the peak magnification is 
$A_\max\ga 100$, just from comparison
of the peak and baseline flux ($I_{\rm base}\sim 19.0$, not shown).
This implies that the strong perturbations
are due to a central caustic probing normalized lens-source separations
$u\ll 1$ (since $A\sim u^{-1}$).

The OGLE data and Turitea/Ellinbank data show
clear caustic exit (B) and entry (D), respectively.
In between, there is a clear ``bump'' (C), which is characteristic of a cusp
crossing or cusp approach.  These features already tell us that
there are three cusps on one side of the primary lens, which
by itself implies that the caustic has one of the three
topologies shown in Figure 2: resonant (6-sided) caustic for
lens-companion separation $s>1$ (normalized to $\theta_\e$),
resonant caustic for $s<1$ (both with $|s-1|\ll 1$), or non-resonant
(4-sided) caustic (same for $s>1$ or $s<1$).  The fact that the
lightcurve is roughly symmetric about the central bump shows that
the trajectory passed roughly perpendicular to the primary-companion axis.
See Figure 2.  The caustic exit,
traced by OGLE, CTIO, and MOA B\&C data (E) then
clearly favors the resonant topologies over the central caustic.
This is because the time from the bump midpoint (C) to the caustic 
re-entrance (D)
is very short compared to the time from re-entrance (D) to final exit (E).
This
accords well with the resonant topologies but clearly contradicts
the non-resonant topology.  The fact that the caustic is both small
and resonant already tells us that this is a planetary event $q\la 0.01$.
That is, roughly equal mass binaries can produce small caustics if
$s\gg 1$ or $s\ll 1$, but in this case they are non-resonant, roughly 
equilateral, 4-sided caustics.

The fact that resonant caustic has a ``close''($s<1$) rather than 
``wide'' ($s>1$) topology is not easily discerned by eye, but can be 
understood qualitatively with the help of Figures 1 and 2.
First, the fact that the Turitea-Ellinbank caustic entrance (D) is slightly lower
than the OGLE caustic exit (B) in Figure~\ref{fig:one} tells us that
the source was further from the center of magnification.  Therefore,
it was not moving quite perpendicular to the star-planet axis, but rather
slightly to the left in Figure 2.
Note that for the wide topology, the top and bottom cusps are to the right of the
central cusp, while for the close topology they are to the left.  The
reason for this difference is deeply rooted in the nature of planetary
lensing.  For $s>1$, as $s$ further increases, these outward cusps will
move further to the right (toward the position of the planet) and eventually
break off to form a quadrilateral caustic on the same side as the planet.
By contrast, for $s<1$ as $s$ further decreases, the outward cusps will
move to the left and eventually break off to form two triangular caustics
on the opposite side of the planet.  This asymmetry reflects the fact that
a point lens has two images, one outside the Einstein ring on the same
side as the source, and one inside the Einstein ring on the opposite
side.  In the planetary ($q\ll 1$) limit, the planet betrays its presence
by perturbing one of these two images.
A planet outside the Einstein ring ($s>1$) must perturb the first
and hence the image must be on the same side as the planet, 
while for $s<1$ it must be on the opposite side of the 
planet\footnote{In the strict planetary limit, $q\rightarrow 0$,
the outer cusps are perfectly aligned with the central cusp at $s=1$,
but for finite $q$ (and $s=1$) the outer cusps lie increasingly to the
right of the central cusp and only ``pass'' it at roughly
$s_{\rm vertical}\sim 1 - 0.55q$.  In the present case, the planetary
limit applies, but in general one should be aware of this effect.}.

Hence, the ``tilt'' of the trajectory implies that it is headed near
the outward
cusp (E) in the close topology, and away from the cusp in the wide topology.
This implies a much longer delay until the caustic exit for the close
topology.  See Figure 2.  The model in Figure 1
shows that the time from first caustic entrance (86.2 (A)) to central cusp
passage (86.9, (C)) is substantially shorter than from central cusp (C) to
second caustic exit (87.8, (E)), consistent with a close topology.  
Unfortunately, the coverage of the
first caustic entrance (A) is not complete enough to securely identify it without
the aid of a model.  Hence, full modeling is required to finally determine
that $s<1$ is the correct topology.  In fact the wide topology is
rejected at $\Delta\chi^2\sim 6000$.

\subsection{Mathematical Model}

We model the light curve using Stokes' Method \citep{gg97} for highly
perturbed regions of the lightcurve and hexadecapole \citep{pejcha09,gould08}
for the moderately perturbed regions.  To account for limb darkening,
we use a linear limb-darkening law and model the source as $n=10$ annuli for
the Stokes integration (after testing that $n=20$ yields essentially
the same results) and use the prescription of \citet{gould08} for
hexadecapole.  We derive coefficients 
$(u_V,u_R,u_I)$ = (0.6504,0.5756,0.4955) from \citet{claret00}
using $T_\eff=6125$K and $\log g=4.0$, based on the measured color
$(V-I)_0=0.585$ and the inferred source radius 
$R_*\simeq R_0\theta_* = 1.25 R_\odot$ (see Section 4).
These correspond to $(\Gamma_V,\Gamma_R,\Gamma_I)$ =
(0.554,0.475,0.396).

\begin{deluxetable}{crr}             
\tablecaption{Microlens Parameters}                      
\tablewidth{0pt}                     
\tablehead{ \colhead{Parameter}      
           & \colhead{Value}         
           & \colhead{Error}}        
\startdata                           
$t_0-6386$                                &      0.9199 &       0.0009 \\
$u_0\times 10^3$                          &       13.23 &         0.04 \\
$t_{\rm E}$(days)                         &       13.23 &         0.05 \\
$\rho\times 10^3$                         &        1.54 &         0.01 \\
$s$                                       &      0.9857 &       0.0001 \\
$q\times 10^3$                            &        3.01 &         0.02 \\
$\alpha$(deg)                             &       261.5 &          0.2 \\
$I_s$(OGLE)                               &      19.205 &        0.003 \\
$f_b/f_s$(OGLE)                           &       0.175 &        0.004 \\
$\theta_{\rm E}$ (mas)$^{\rm A}$          &       0.456 &        0.003 \\
$\mu_{\rm geo}$ (mas yr$^{-1}$)$^{\rm A}$ &       12.57 &         0.08 
\enddata                             
\tablecomments{(A) Error shown is from lightcurve only.  Additional 7\%     
systematic error is discussed in text.}                                
\label{tab:one}                    
\end{deluxetable}   

Table~\ref{tab:one} shows the resulting microlensing fit parameters.  Here,
$(u_0,t_0,t_\e)$ are the parameters of the underlying event, respectively
the impact parameter (normalized to $\theta_\e$) of the source trajectory
relative to the
center of magnification of the lens system, the time of closest approach,
and the Einstein radius crossing time.  The planet parameters are 
$(s,q,\alpha)$, respectively the normalized planet-host separation,
the planet-host mass ratio, and the angle of the planet-host axis
relative to the source trajectory.  Finally, $\rho\equiv\theta_*/\theta_\e$ 
is the normalized source radius, while $f_s$ and $f_b$ are the fluxes
due to the lensed source and the unlensed blended light in the aperture,
respectively.

\section{Physical Parameters}

\subsection{Angular Scale}

The source crossing time, $t_* = \rho t_\e = 29\,$min, is exceptionally
short for microlensing events, indicating either a high proper motion
or very small source.  From regression of CTIO-SMARTS $V/I$ data, 
we derive a model-independent instrumental source color 
$(V-I)_{s,\rm inst}=-0.39$
and from the model we obtain $I_{s,\rm inst}=19.46$.  We measure the 
instrumental 
position of the clump as $(V-I,I)_{\rm cl,inst} = (0.07,15.94)$ and adopt
$(V-I,I)_{0,\rm cl}=(1.06,14.39)$ from \citet{bensby13} and \citet{nataf13},
and so derive $(V-I,I)_{0,s}=(0.60,18.04)$.  Using a similar procedure
for OGLE $V/I$ data, we infer $(V-I,I)_{0,s}=(0.57,17.91)$.  We adopt
the average of these.  Using the $VIK$ color-color
relations of \citet{bb88} and the color/surface-brightness relations
of \citet{kervella04} we 
find $\theta_*=0.704\mu$as.  We then find,
\begin{equation}
\mu_\geo = {\theta_*\over t_*} = 12.5\,\masyr;
\qquad
\theta_\e = {\theta_*\over \rho} = 0.45\,{\rm mas},
\label{eqn:thetae}
\end{equation}
where $\mu_\geo$ is the instantaneous lens-source relative proper motion
in the geocentric frame at the peak of the event.
The principal errors in these quantities derive from the color-magnitude
offset of the source from the clump, rather than the microlensing
fit parameters $(t_\e,\rho,I_s)$.  The color error is known from spectroscopy 
of a microlensing sub-sample to be 0.05 mag 
\citep{bensby13}\footnote{The overall scatter between $(V-I)_0$ as
inferred from microlensing color (i.e., method used here) and
from spectroscopy in the \citet{bensby13} sample is about 0.07 mag.
However, as those authors discuss, this scatter is dominated by
relatively red sources, for which the most likely cause is errors
in the photospheric models.  Hence, it is more appropriate to adopt a smaller
uncertainty for bluer sources like MOA-2013-BLG-220.
In addition, roughly 0.03 mag of the scatter (in quadrature)
is caused by measurement error in the spectroscopic temperature.}.  
We estimate the magnitude error, which is basically due to uncertainty
in centroiding the clump, to be 0.10 mag.  This value is consistent
with general practice and also with the difference between the 
CTIO-SMARTS-based and OGLE-based determinations.  In principle, one might
adopt an error that is smaller by $\sqrt{2}$ due to averaging two
determinations, but we conservatively decline to do so.
Combined these yield a fractional error in $\theta_*$
of 7\%, which then propagates directly to uncertainties in 
$\mu$ and $\theta_\e$.

\subsection{Mass-Distance Constraint}

This measurement of $\theta_{\rm E}$ directly relates the lens
mass and distance via Equation~(\ref{eqn:mpirel2})
\begin{equation}
M\pi_\rel = {\theta_\e^2\over\kappa} = 0.025\,M_\odot\,{\rm mas}.
\label{eqn:mpirel3}
\end{equation}
If the lens were in or near the bulge, then
$M\simeq 1.7\,M_\odot(\kpc/D_{LS})$, where $D_{LS}=D_S-D_L$ is the distance
between the lens and source.  However, unless the lens is a black hole or
neutron star, it must be $M<1\,M_\odot$.  This is because the source
has $M_I\sim 3.4$ and the lens (which is closer) must be at least $\sim 1$
mag dimmer or it would be seen in the blended light.  Therefore, the
lens must be more than 1.7 kpc in front of the source 
($D_L<6.5\,\kpc$, $M<0.77\,M_\odot$)
and so almost
certainly in the Galactic disk.

Because Turitea and Ellinbank observatories are
separated by about 2500 km and both cover the caustic entrance,
which is a very sharp feature in the lightcurve, it is in principle
possible to measure the ``microlens parallax'' vector 
$\bpi_\e=(\pi_{\e,\rm North},\pi_{\e,\rm East})$
\citep{gould04,hardy95,ob07224}
from the lightcurve differences (after taking account of limb-darkening
differences).  Even a very weak constraint would be of some interest.
For example if $\pi_\e<3$, then the lens-source relative parallax
would be $\pi_\rel=\theta_\e\pi_\e < 1.3$, meaning that the lens would
be further than 700 pc.  Unfortunately, we find no meaningful constraints
at the 3 sigma level.

\section{Source-Lens Relative Proper Motion}

The high proper motion in Equation~(\ref{eqn:thetae}) should
be compared to our expectation for
typical lens-source relative proper motions.  
A typical proper motion for a lens in the bulge is
 $\mu\sim 4\,\masyr$
and for a disk lens,
$\mu\sim \mu_{\rm sgrA*}=6.4\,\masyr$.  The former is
due to typical 1-D lens and source dispersions of $\sim 100\,\kms$
at Galactocentric distance
$R_0\sim 8.2\,\kpc$ while the latter is due to the Sun and lens partaking
of the same approximately flat Galactic rotation curve.  

There are
two basic ways to produce higher proper motions: either lenses and/or
sources that are moving very fast relative to their populations, or
nearby lenses.  In the latter case, we expect that the typical
peculiar motions of stars relative to their local LSR, $v_{\rm pec}\sim 30\,\kms$
will add as a more-or-less randomly oriented vector 
$\Delta\mu = v_{\rm pec}/D_L\sim 6\,\masyr (\kpc/D_L)$.  Hence, if this is
the cause, one expects that the lens will be within a few kpc.

While the MOA-2013-BLG-220 ``source star'' (actually, ``baseline object'')
 mostly fell between chips in the 
OGLE-IV survey, it was observed 1298 times between 2001 and 2009 by OGLE-III.
We are therefore able to measure its proper motion (relative to a frame
of Galactic bulge stars) 
\begin{equation}
\bmu_{\rm base} = (\mu_\ell,\mu_b)= (-5.6,+1.9)\,\masyr
\label{eqn:sourcepm}
\end{equation}
with an error of $1.2\,\masyr$ in each direction.  That is, the 
baseline object is moving almost directly opposite to the
direction of Galactic rotation.  

However, the light in the baseline object may also include light from
stars other than the source. Hence, while the source dominates the
light from the baseline object, a minority of the light may be coming
from the lens or some other, unrelated object.
Nominally, our
measurement indicates that 15\% of the baseline flux, $f_{\rm base} =
f_{\rm s} + f_{\rm b}$, comes from the blended light ($f_{\rm b}$)
rather than from the source ($f_{\rm s}$). However, this may be
because the estimate of the baseline flux is itself the result of
crowded-field photometry (DoPHOT, \citealt{dophot}),
rather than difference imaging and so is subject to
considerable uncertainty.  If this $(f_{\rm b}/f_{\rm base})$ estimate were by chance
perfectly accurate and if, for example, the blended light is not
moving in the bulge frame, then 
$\mu_s = \mu_{\rm base} f_{\rm base}/f_s =7.2\,\masyr$.  
Hence, the $\bmu_{\rm base}$
measurement
is a strong qualitative indication of source motion, though
not a direct determination
of it.  Nevertheless, if we adopt $\bmu_{\rm s}\sim \bmu_{\rm base}$ as a proxy, and
consider typical disk lens motion (relative to the bulge frame)
$\bmu\sim (\mu_{\rm sgrA*},0)$, then this would predict 
$\mu_\rel\sim 5.6+6.4\sim 12\,\masyr$.  That is, a disk lens combined
with the observed retrograde motion of the baseline object, would
naturally reproduce the relative proper motion derived from the lightcurve
analysis.

Since the mass-distance constraint (Equation (\ref{eqn:mpirel3})) combined
with upper limits on the lens flux already imply $D_{LS}\ga 1.7\,\kpc$,
and the measurements of $\mu_\rel$ and $\bmu_{\rm base}$ favor disk-lens
kinematics, we conclude that the lens is almost certainly in the disk.

\section{Future Mass Determination}

Because of the wide range of possible lens masses, we do not try to give
a ``best estimate''.  One could in principle attempt a Bayesian analysis
based on a Galactic model, but this would require either assessing
(or more likely ignoring) the priors on, e.g., an $M=0.05\,M_\odot$ 
brown dwarf at $D_L = 1.6\,\kpc$
having an $m=50\,M_\oplus$ planet at projected separation
$r_\perp = 0.7\,\au$ relative to an $M=0.8\,M_\odot$  star at $D_L = 6.5\,\kpc$
having an
$m=3.8\,M_{\rm Jup}$ planet at $r_\perp=3\,\au$.  Since this prior 
on relative planetary frequency is completely unconstrained, a
Bayesian posterior estimate of the primary mass would essentially
reflect the prior on the primary mass, and therefore would not be
informative.

Rather we focus on how the mass can be determined.  
First, one could measure the excess light (above the known flux
from the source) in high-resolution images taken immediately.
One limitation of this approach is that if the lens were even
half as bright as the source then it would have been noticed already,
and if it were less than a tenth as bright, it would be difficult
to unambiguously detect.  In addition, one must be concerned about
non-lens stars generating the excess light, which could be either
a random star along the line of sight or a companion
to either the source or lens (e.g., \citealt{mb11293B}).

However, the high lens-source proper motion almost
guarantees that this issue can be resolved by high-resolution imaging
when the source and lens are separated by $\sim 100\,$mas, roughly
8 years after the event, i.e., 2021.  We note that while 
Batista et al.\ (2014, in preparation) have detected a planetary host
lens separated from its source star by only 60 mas
using Keck adaptive optics, the source and lens fluxes were comparable in
that case.  In the present case, a lens at the bottom of the main sequence
would be $H\sim 24$ and therefore roughly 100 times fainter than the
source.  Hence, to detect (or rule out) such a lens requires significantly
greater separation.  If the lens is not seen at that point, it must be ``dark'',
which in practice means either a white dwarf at distance $D_L\sim 6\,\kpc$
or a brown dwarf at $D_L\la 2.3\,\kpc$.

Note also that 
$\bmu_\hel-\bmu_\geo={\bf v}_{\oplus,\perp}\pi_\rel/\au$
where ${\bf v}_{\oplus,\perp}=(3.2,6.6)\,\kms$ is the velocity of
Earth projected on the plane of the sky in equatorial coordinates (N,E).
Therefore, if the lens is on the main sequence (and hence visible),
then by Equation~(\ref{eqn:mpirel}) $\pi_\rel<0.3$, so that
this correction to the proper motion is less than $0.5\,\masyr$.

\section{Characterization Using Only Followup Data}

In the era of second generation microlensing surveys, the relationship
between survey and follow-up data is becoming more complex.  From the
standpoint of designing and implementing future strategies, it is important
to establish the conditions under which planets can be characterized
by survey-only data and by followup-only data, particularly for 
high-magnification events.  For example, \citet{mb11293} showed that
although MOA-2011-BLG-293Lb was in practice discovered in followup data,
it could have been well-characterized by survey data alone.  Although
\citet{mb11293} did not explicitly address this question, the converse
is not true: without survey data it would have been impossible to
even approximately measure the event timescale, $t_\e$ for
that event.  Similarly, 
\citet{ob120406b} investigated how well \citet{ob120406} had been
able to characterize OGLE-2012-BLG-0406Lb based on survey-only data.

Here we ask the opposite question: how well can MOA-2013-BLG-220 be
characterized by followup-only data?  This may seem like a Scholastic
question, given that there are always survey data (since otherwise
the event could not be ``followed up'').  But already at present, of order
100 square degrees are monitored at low cadence ($\la 0.5\,\rm day^{-1}$)
and in the future many hundreds of square degrees may be monitored
at even lower cadence (e.g., \citealt{gouldlsst}).  These survey data may
be too thin to provide any significant constraint on event characterization,
and therefore may be usefully approximated as being absent.

Because survey telescopes can switch into followup mode, it is not
necessarily trivial to determine which observations should be assigned
to each category.  For example, in the case of MOA-2011-BLG-293, the cadence
of OGLE data was greatly increased on the night of the anomaly alert 
relative to the normal cadence of $\sim 3\,\rm hr^{-1}$.  \citet{mb11293}
therefore approximated the ``survey portion'' as a subset following
this cadence.

In the present case, we are fortunate that OGLE had to repoint the
telescope in order to guarantee placement of the event on a chip.
Thus we can reconstruct precisely, from the image
headers, which OGLE observations
are ``followup''.  These turn out to be all observations $6386<t<6391$
and no others.

For MOA, the situation is more complicated.  The cadence is much
higher on the night of the second caustic entrance, which was
in specific response to the alert.  The decision to point the MOA 
telescope at this target for the single point on the cusp
approach, which was observed earlier the same night during a momentary
improvement in observing conditions, may have been strongly influenced
by the alert as well.  However, all of these points, whether considered
``survey'' or ``followup'', are closely matched by observations from
the MOA B\&C followup telescope operating from the same site, and
all the caustic-entrance points are covered by Turitea and Ellinbank as
well.  The one place where MOA data uniquely constrain the event is
on the first caustic entrance.  However, these took place before the
alert and so are unquestionably ``survey data''.
Hence, inclusion or exclusion of MOA followup data does not
materially affect the followup-only fit.  Therefore, for simplicity and
to be conservative, we eliminate all MOA data from the followup-only
analysis.

\begin{deluxetable}{crr}             
\tablecaption{Parameters from Followup-only Data}                      
\tablewidth{0pt}                     
\tablehead{ \colhead{Parameter}      
           & \colhead{Value}         
           & \colhead{Error}}        
\startdata                           
$t_0-6386$                                &      0.9197 &       0.0007 \\
$u_0\times 10^3$                          &       13.20 &         0.05 \\
$t_{\rm E}$(days)                         &       13.27 &         0.07 \\
$\rho\times 10^3$                         &        1.54 &         0.01 \\
$s$                                       &      0.9857 &       0.0001 \\
$q\times 10^3$                            &        3.00 &         0.02 \\
$\alpha$(deg)                             &       261.5 &          0.2 \\
$I_s$(OGLE)                               &      19.205 &        0.003 \\
$f_b/f_s$(OGLE)                           &       0.112 &        0.035 \\
$\theta_{\rm E}$ (mas)$^{\rm A}$          &       0.456 &        0.003 \\
$\mu_{\rm geo}$ (mas yr$^{-1}$)$^{\rm A}$ &       12.52 &         0.08 
\enddata                             
\tablecomments{(A) Error shown is from lightcurve only.  Additional 7\%     
systematic error is discussed in   text.}
\label{tab:two}                    
\end{deluxetable}   

Table~\ref{tab:two}, shows the fit parameters resulting from
models with only the followup data included.
Comparison with Table~\ref{tab:one} shows
that all the parameters are the same within errors and that the
error bars themselves are only slightly larger.

\section{Discussion}

This is the third event published in the era of survey-only microlensing
planet detections (i.e., post-2010)
in which follow-up observations play a decisive role,
the other two being MOA-2011-BLG-293 and OGLE-2012-BLG-0026.  We review
the common features of these, which may be indicative of the role of
follow-up observations in this era.

First, all three events were high-magnification.  In such events, the
planetary perturbations are often of shorter duration than in low-magnification
events because the caustic is smaller.  In addition,
in contrast to low-magnification events, both the event
characteristics and the planetary perturbation must be fit from the
same peak data.  Hence, these events
benefit from the denser coverage that is made possible with follow-up.
Also, the existence of multiple sites permits aggressive action in the
face of weather or equipment problems at any given site.

Two of the events were in the wings of the season (early March and April,
respectively), when pure-survey coverage is intrinsically limited and,
again, aggressive multi-site follow-up can compensate for the shortness
of the observing night.

Finally, all three events were covered by CTIO-SMARTS, where $H$-band data
are automatically taken using the dichroic ANDICAM camera 
\citep{depoy03}.  As in the present case, the $H$-band data are usually
not incorporated into the lightcurve analysis because they have
much lower signal-to-noise ratio than the contemporaneous $I/V$ data.
However, these data proved crucial to the final interpretation
of OGLE-2011-BLG-293Lb based on AO $H$-band observations taken a year 
after the event \citep{mb11293B}, and are likely to prove crucial
for MOA-2013-BLG-220Lb as well.

MOA issued its alert relatively early in the season when observations
were possible only $\sim 5\,$ hours per night.  Given their $1.5\,\rm
hr^{-1}$ cadence and variable conditions, this enabled 3--5 points per
night.  On the night of the $\mu$FUN alert, the event was already
bright enough, and hence the photometric errors small enough, that the
continued rise during the night was visible, which aided considerably
in making the determination that the event was headed toward high
magnification and so should be intensively observed.  Hence, this is
{\it not} an example of a call for high-cadence observations being
issued on the basis of truly low-cadence survey data.  For that,
software of the type being developed by RoboNet would be necessary.
Nevertheless, the fact that the planet
could be characterized by followup observations alone shows that
alerts derived from such low-cadence survey data (combined with
automated moderate-cadence ``patrols'') can yield planet detections
that are fully characterized.  This opens the prospect for
complementing high-cadence surveys with very wide low-cadence surveys,
when the latter are coupled to aggressive followup.  Such a strategy
could be employed over very wide ($\sim 6000\,\rm deg^2$) areas using
LSST if that project does not exclude the inner two quadrants of the
Galaxy \citep{gouldlsst}.

\section{Conclusions}

We have presented the discovery of a planet with relatively high
mass ratio $q=0.0030$, i.e., three times that of Jupiter and the Sun.
The underlying event has a relatively high proper motion, $12.5\,\masyr$.
Two lines of argument show the planet lies in the Galactic disk.
First, the measured Einstein radius $\theta_\e=0.45\,$mas, together
with an upper limit on the lens flux, implies that the lens lies
at least 1.7 kpc in front of the source.  Second, the source (actually,
``baseline object'') proper motion is $\sim 6\,\masyr$ counter to 
Galactic rotation, implying that typical disk-lens motion of $\sim 6.5\,\masyr$
would naturally produce the observed lens-source relative proper motion.
The actual lens mass and distance can be measured in
the short term by looking for excess flux at the
position of the lens in {\it HST} or ground-based AO observations provided
it is at least 10\% of the source brightness and otherwise by $\sim 2021$,
i.e., once the lens and source have moved far enough apart to be separately
resolved.

\acknowledgments
Work by J.C.\ Yee is supported in part by a Distinguished University Fellowship
from The Ohio State University and in part under contract with the California 
Institute of Technology (Caltech) funded by NASA through the Sagan Fellowship 
Program.
Work by CH was supported by Creative Research Initiative Program
(2009-0081561) of National Research Foundation of Korea.
Work by AG and BSG was supported by NSF grant AST 1103471.
Work by AG, BSG, and RWP was supported by NASA grant NNX12AB99G 
TS acknowledges the support from the grant JSPS23340044 and JSPS24253004.
The OGLE project has received funding from the European Research Council
under the European Community's Seventh Framework Programme
(FP7/2007-2013) / ERC grant agreement no. 246678 to AU.
This publication was made possible by NPRP grant 
X-019-1-006 from the Qatar National Research Fund (a member of Qatar
Foundation).

\end{document}